# Analytical Charge-Control Model for DH-HEMT

Petrica I. Cristea

*University of Bucharest, Department of Physics, Bucharest-Magurele MG11,* Romania
*E-mail: pcristea@physicist.net*

**Abstract. -** A new and simple analytical charge-control model of the two-dimensional electron gas of a double-heterojunction $Al_{Y_1}Ga_{1-Y_1}As/GaAs/Al_{Y_2}Ga_{1-Y_2}As$ HEMT is described. The analytical calculation is based on an adapted version of the triangular well approximation. The charge-control equation accounts for a variable average distance of the 2DEG in the under-gate region. An analytical relation is introduced to describe its dependence on 2DEG concentration. It is found that, despite its simplicity, the charge-control model gives an accurate description of the device operation for a wide range of physical parameters ($Y_{1,2}=0.25$-$0.36$, $d_1=100$-$450A$, $d_{i1,i2}=10$-$100A$, $d_w=250$-$350A$), and gate voltages $V_g = -3.5$-$1V$. The validity of the analytical model is supported by the calculated results of the self-consistent quantum mechanical model and agrees with previously reported theoretical and experimental data. The influence of some physical parameters, such as layer thickness and aluminum composition, on the device performance is also discussed.

1. Introduction

In recent years, the fabrication of nanoscale devices has become a standard technology. Structures such as High-Electron-Mobility-Transistors (HEMTs or MODFETs) and resonant devices have an increasing importance for high-speed digital applications and far-infrared detectors. Their theory and technology were continuously and substantially improved. A detailed presentation of MODFET principles, fabrication techniques and applications is given in ref. [86DRU]. For a single-interface HEMT structure, the two-dimensional electron gas (2DEG) achieves a density of $5$-$7 \times 10^{15}$ $m^{-2}$ and a mobility of $0.6$ $m^2(Vs)^{-1}$, as determined from Hall measurements at 300K. The latter sheet density is too small for power applications. If the contribution of two or more heterojunction channels is added, the electron density is increased to $1.5$-$2 \times 10^{16}$ $m^{-2}$, without excessive donor densities [85GUP, 86HIK, 86SAU, 88WAN, 88CAZ]. For a single-interface, the 2DEG density can be related with the Fermi level position by using self-consistent calculation results [84STE], or using analytical relations [85MOL, 88SHE]. In a double-heterojunction (DH) structure, the top and bottom 2D electron gases interact. In some cases, this interaction is a source of significant contribution to energy. For a large quantum well (QW), exceeding in thickness 250 A, the coupling term adds only a small contribution to energy and generally can be treated as a perturbation. Its effect can, with some cautions, be measured by the overlap integral calculated with the wave functions of the top and bottom channel. For narrow QW systems, such a treatment is not a priori legitimate and to obtain an accurate prediction of the total 2DEG density, at a given gate voltage, self-consistent numerical models must be used [85INO, 87JAF].

In this letter, a simple analytical charge-control model for a DH-HEMT has been developed for use of analytic C-V and I-V modeling. We show that our model well fits self-consistent calculations and allows an easy and fast calculation of the device performance.

## 2. The model
### 2.1 Structure and physical operation

Fig.1a shows a schematic and self-explanatory cross-section of the DH-HEMT structure. The under-gate region of the conduction-band energy diagram for this structure is shown in Fig.1b. $E_{F1}$ and $E_{F2}$ are the positions of the flat-Fermi level, measured from the bottom of the conduction band in the QW region, at the top and bottom interface, respectively. The voltage $V_g$ is applied to gate electrode with respect to source electrode. The permitivity of the n-doped ($N_{D\alpha}$) $Al_{Y\alpha}Ga_{1-Y\alpha}As$ layer is $\varepsilon_\alpha$, ($\alpha=1$ or 2, and 1 stands for top and 2 for bottom). The quantum-well region (GaAs), with the permitivity $\varepsilon$, is undoped (or with a background concentration of acceptor impurities, $N_A$). To further reduce the coulombian coupling between the ionized donors and the 2DEG channels, thin undoped $Al_{Y\alpha}Ga_{1-Y\alpha}As$ coulombian spacers are inserted between the QW region and n-AlGaAs layers.

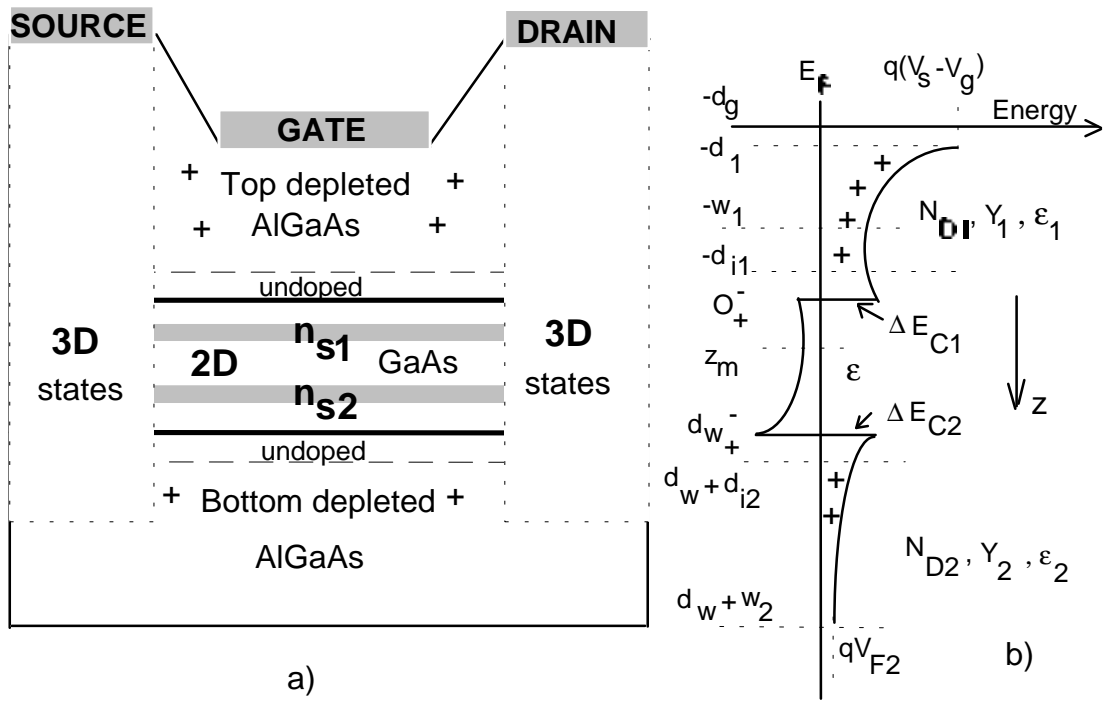

Fig.1 Schematic cross-section of the DH-HEMT structure used in modeling (a), and diagram of the conduction-band edge under external gate voltage $V_g$ (b)

For a better understanding of the structure operation we studied two representative situations:

1. In the top AlGaAs layer, along z-axis of the structure, the dependence of the conduction band $E_c(z)$ shows a "minimum" located at $-w_1$. Then, the law of charge conservation tells us that, in the QW region, $E_c(z)$ must exhibits a "maximum" located at $z_m$. Within this picture, the expressions for the electric field $F_\alpha$ at the top and bottom interface, in the QW region, are respectively:

$$F_1 = \frac{q}{\varepsilon}(n_{s1} + \theta_1 z_m N_A) \tag{1a}$$

$$F_2 = -\frac{q}{\varepsilon}\left[n_{s2} + \theta_2(d_w - z_m)N_A\right] \tag{1b}$$

where, *by definition*, $n_{s\alpha}$ is the top or bottom electron sheet density, and the ratio $\theta_\alpha \in (0,1)$ describes the ionization of the acceptor impurities. To obtain these ratios and $z_m$ it is necessary to have a simultaneous solution of Schrödinger's and Poisson's equations. We note that (1a) and (1b) are only approximations, since the true electrostatic potential cannot have any extreme (Earnshaw's theorem states that a charge, acted on by electric forces only, cannot rest in stable equilibrium in an electric field). We have obtained these equations by neglecting, in applying the Gauss' law, the electric flux due to the field along x and y-axis. The model is applicable to the situation of a 2DEG channel with no source-drain voltage and for regions away the source and drain contacts on 2DEG. These are transitions regions were an exchange between 3D (contacts) and 2D (QW) electronic states occurs. It is still not clear how the contact between a 3D system and a 2D system looks on a microscopic level. In the following we suppose that (1a) and (1b) are valid approximations. By changing the gate voltage, the electrostatic equilibrium along the structure changes and the total sheet density, $n_s = n_{s1} + n_{s2}$, is increased or lowered. To recover the equilibrium, the length $w_1 - d_{i1}$ of the top AlGaAs depleted region will be higher or lower, respectively (see also Fig.1b). As long all donors remain fully ionized and the background doping is neglected, the variation of the top sheet density, $\Delta n_{s1}$, is:

$$\Delta n_{s1} \cong N_{D1}\Delta w_1 \tag{2}$$

By applying the Gauss' law for the region between $z = -d_g$ and $z = -w_1$, within above approximations, we have:

$$\Delta n_s = \Delta(n_{s1} + n_{s2}) \cong \Delta n_{s1} \tag{3}$$

As a result, when $n_{s1} \neq 0$, the coupling between the bottom channel and the gate voltage appears to be very weak, and a plateau in the electron density $n_{s2}$ is expected. If the gate voltage is near $V_{g,th1}$, namely the threshold value for the top channel, the density $n_{s1}$ in the top channel is negligible and the electric field at the top interface $F_1$ vanishes. These observations are all in good agreement with the calculated band-diagram [85INO].

   2. When $V_g \leq V_{g,th1}$, and approaches $V_{g,th}$ (the absolute threshold voltage), at second interface the Fermi level moves to the bottom of the QW. The sheet density $n_{s2}$ gradually decreases and the 2D system becomes highly confined, resulting an increase of the subbands energies $E_i$ (i=0,1...). For a given temperature $T$, the 2DEG density and the electric field at the bottom interface are given by the well-accepted expressions, [82DEL]:

$$n_i = Dk_BT \ln\left[\exp\left(\frac{E_{F2} - E_i}{k_BT}\right) + 1\right] \tag{4a}$$

$$n_s = \sum_i n_i \tag{4b}$$

and [(A6), Appendix I]:

$$F_2 = -\frac{qN_{D2}}{\varepsilon}\left[\sqrt{\frac{2\varepsilon_2}{qN_{D2}}\left(\frac{\Delta E_{C2} - E_{F2}}{q} - V_{F2}\right) + d_{i2}^2} - d_{i2}\right] \quad (5)$$

where $D = \frac{m^*}{\pi\hbar^2}$ is the density of states in a 2D system, including a factor of two for spin degeneracy, and $V_{F2} = (E_C - E_F)/q + V_{FD}$. In this latter expression, the second term $V_{FD}$ is a correction voltage for Fermi-Dirac rather than Boltzmann statistics, and $E_C$-$E_F$ refers to the degenerate semiconductor case, away the heterojunction [83LEE]. If $E_{F2} \cong 0$, the triangular potential well approximation (TWA) yields to a quite satisfactory solution for the 2DEG density as a function of $E_{F2}$ (very good agreement is seen between the analytical and our numerical self-consistent data). Accordingly, in the envelope-function approximation (EFA), the Schrödinger equation leads to the well-known Airy equation, with energies [67STE]:

$$E_i = \gamma_i F_2^{2/3} \quad (6)$$

where, for GaAs and in units of $(eV) \times (m/V)^{2/3}$, the coefficients $\gamma_i$ are:

$$\begin{aligned}\gamma_0 &= 1.968 \times 10^{-6} \\ \gamma_1 &= 3.442 \times 10^{-6} \\ \gamma_2 &= 4.648 \times 10^{-6}\end{aligned} \quad (7)$$

The energies, as given by (7), are measured from the bottom of the conduction band in GaAs, at the bottom interface.

### 2.2 Fermi level at the TOP and BOTTOM interface

If the z-component of the current density is neglected, the flat-Fermi level approximation well works. To obtain the charge-control equation, we seek for a relationship between the positions $E_{F1}$ and $E_{F2}$ of the Fermi level at the top and bottom interfaces. Recall that $E_{F_i}$ are measured with respect to the bottom of the conduction band edge in the QW, at the top and bottom interfaces. The Poisson's equation for the GaAs potential well is:

$$\frac{d^2\varphi}{dz^2} = \frac{qn_s}{\varepsilon}\sum_i f_i \psi_i^2(z) + \frac{qN_A(z)}{\varepsilon} \quad (8)$$

where $\psi_i$ is the wave functions for the QW quantized motion along z-axis in the i-th subband, and $f_i = n_i/n_s$ is the occupation factor. Each $\psi_i(z)$ must satisfy Schrödinger's equation in the effective mass approximation:

$$\frac{\hbar^2}{2m^*}\frac{d^2\psi_i}{dz^2}+[E_i-V(z)]\psi_i=0 \tag{9}$$

By multiplying both sides in (8) with $z$, and integrating between $0^+$ and $d_w^-$, one obtains:

$$E_{F1}-E_{F2}=qF_2d_w+\frac{q^2 n_s}{\varepsilon}Z_{av}+\frac{q^2\theta N_A}{2\varepsilon}d_w^2 \tag{10}$$

where:

$$Z_{av}=\sum_i f_i \int_{0^+}^{d_w^-} z\psi_i^2 dz \tag{11}$$

is the average distance of the 2-DEG from the top interface, and $\theta\in(0,1)$ (average fraction of ionized acceptors). The average distance $Z_{av}$ is obviously a function of 2DEG density. Its accurate determination requires a self-consistent numerical procedure. Thus, equation (10) tells us that the difference in the energy of the bottom of the quantum wells, at the two interfaces, is a sum of contributions arising from the field $F_2$, the field due to 2D electrons (second term in the right-hand side), and finally, the field of the background impurities. In almost all practical situations, the latter contribution can be neglected without making significant errors in modeling.

The total 2DEG sheet density in the GaAs layer is given by:

$$n_s=\frac{\varepsilon}{q}(F_1-F_2)-\theta\, d_w N_A \tag{12}$$

and (10) reads:

$$E_{F1}=E_{F2}+qF_1d_w-\frac{q^2 n_s}{\varepsilon}(d_w-Z_{av})-\frac{q^2\theta N_A}{2\varepsilon}d_w^2 \tag{13}$$

*2.3 Charge-control equation $n_s=n_s(V_g)$*

Using the results of the Appendix I, (10) and (12), after a straightforward algebra we obtain the folowing charge-control equation:

$$n_s=\frac{\varepsilon_1}{q(d_1+\Delta d)}\left(V_g-V_S+\frac{\Delta E_{C1}-E_{F2}}{q}+\frac{qN_{D2}d_{eff}^2}{\varepsilon_1}\right) \tag{14}$$

where $d_{eff}$ is defined as:

$$d_{eff}^2=(w_2-d_{i2})\left(\frac{\varepsilon_1}{\varepsilon}d_w+d_1\right)+\frac{N_{D1}}{2N_{D2}}(d_1-d_{i1})^2-\left(\frac{\varepsilon_1}{2\varepsilon}d_w+d_1\right)\frac{\theta N_A}{N_{D2}}d_w \tag{15}$$

and:

$$\Delta d = \frac{\varepsilon_1}{\varepsilon} Z_{av} \tag{16}$$

The first term in the righ-hand side of (15) is a function of $E_{F2}$ through $w_2$, the thickness of the depleted region in the bottom AlGaAs layer [$d_{i2}$ included, (A7), Appendix I]:

$$w_2(E_{F2}) = \left[ \frac{2\varepsilon_2}{qN_{D2}} \left( \frac{\Delta E_{C2} - E_{F2}}{q} - V_{F2} \right) + d_{i2}^2 \right]^{1/2} \tag{17}$$

the second and the third terms being constants for a given geometry.

*2.4 Capacitance-Voltage relation*

We consider that the 2DEG contains all the mobile charge, namely $Q_{2DEG}$. The differential capacitance, per unit area, then reads:

$$C_g = \frac{d}{dV_g}\left( -\frac{Q_{2DEG}}{A_{2DEG}} \right) = q\frac{dn_s}{dV_g} \tag{18}$$

By using (14) and (15), after a straightforward algebra, one obtains (Appendix II):

$$C_g = (C_1^{-1} + C_2^{-1})^{-1} \tag{19}$$

where:

$$C_1^{-1} = \frac{d_1}{\varepsilon_1}\left( 1 + \frac{\varepsilon_1}{\varepsilon}\frac{Z_{av}}{d_1} \right) \tag{19a}$$

$$C_2^{-1} = \frac{n_s}{\varepsilon}\frac{dZ_{av}}{dn_s} + \frac{1}{q^2}\left[ 1 + \frac{\varepsilon_2}{\varepsilon_1}\frac{(\varepsilon_1/\varepsilon)d_w + d_1}{w_2} \right]\frac{dE_{F2}}{dn_s} \tag{19b}$$

The capacitance $C_1$ describes the effect of the top depleted donor layer, and includes the contributions arising from $Z_{av}$ and different permitivity of the GaAs quantum well and top AlGaAs layer. The capacitance $C_1$ can be thought of as an effective depleted insulator capacitance (dependent on gate voltage) of the top interface. The first term in the r.h.s. of (19b) is negative and arises from $Z_{av}$ dependence on the sheet density, $n_s$. Increasing $V_g$ causes $Z_{av}$ to decrease and hence increases the device capacitance. Finally, the second term is related with the variation of $E_{F2}$ with $n_s$ and the bottom AlGaAs depleted donor layer effects. When $V_g$ approaches the absolute threshold value $V_{g,th}$, the sheet density is almost negligible and TWA is quite accurate. Accordingly, the average separation $Z_{i,av}$ of the charge carriers in the i-th energetic subband, measured from the top interface, is obtained using some properties of Airy function [72STE]:

$$Z_{i,av} = d_w + (2/3)\gamma_i F_2^{-1/3} \tag{20}$$

where, for GaAs, the coefficients $\gamma_i$ are given by (7). A high electric field is developed at interface (the order of $-2 \times 10^7$ V/m, as can be seen in Fig. 2) and the second term in the r.h.s. of (20) gives only a small contribution (of only 30-40 A) to $Z_{i,av}$, but not entirely negligible when compared to the QW thickness $d_w$ (usually the order of 200 A). Actually, in this voltage range, the capacitance drops to zero and can be calculated almost exactly in the triangular potential well approximation. For a single MODFET interface, such a situation has been analysed in a detailed model by Sadwick and Wang [86SAD].

### 2.5 Description of the Analytical Model

To calculate the electron sheet density $n_s$ as a function of gate voltage $V_g$, the $Z_{av}$ and $E_{F2}$ (or $E_{F1}$) dependencies on $n_s$ must be determined. Rigorously, these relations are derived from a numerical procedure that gives a self-consistent solution for Schrödinger and Poisson's equations, with the proper heterobarriers potential and accounting for electron-electron exchange-correlation and other effects [84STE]. Our numerical results and comparisons with the previously published theoretical and experimental results [88CAZ, 85INO, 87JAF] indicate that a simple algorithm that leads to accurate results avoids many complications:

i) Starting with a negative value for $E_{F2}$ (-0.04 eV in this work) and then increasing it in small steps ($5 \times 10^{-4}$ eV) we calculate at every step the field $F_2$ at the bottom heterojunction with (5), the energies $E_i$, using eqs. (6) and (7) within triangular potential well approximation, and finally the two-dimensional electron sheet concentration $n_s$ with eqs. (4a,b). Only three energetic subbands have been considered in this letter. It should be noted that the starting value $E_{F2}$ must be low enough to ensure an almost vanishing initial concentration in the bottom channel.

ii) Neglecting the background doping $N_A$, in accord with (1a) and (12), we get:

$$n_{s1} = \left( n_s + \frac{\varepsilon}{q} F_2 \right) \cdot \Theta \left( n_s + \frac{\varepsilon}{q} F_2 \right) \qquad (21)$$

where $\Theta$ is the step-function. If the argument under the step-function is positive, both top and bottom QW channels are populated with 2D electrons. Otherwise, $n_{s1} = 0$ while $n_s = n_{s2}$. In this case, the bottom sheet density is simply calculated as $n_{s2} = n_s - n_{s1}$.

However, many questions are still open. Firstly, the triangular well approximation seems to be not adequate when $n_{s2}$ considerably differs from zero. Self-consistent calculations predict that when $n_{s1}$ starts increasing, the bottom sheet density has the order of $6 \times 10^{15} \text{m}^{-2}$. Secondly, as shown in the Sec.2.1, when $n_{s1} \neq 0$, the bottom sheet density $n_{s2}$ should be nearly constant. This means that the field $F_2$ at the bottom interface also is nearly constant (fig. 2), and the position $E_{F2}$ of the Fermi level will correspondingly be pinned. Despite the crude approximations, we have observed that, even for $n_{s1} \neq 0$, by gradually increasing $E_{F2}$ and calculating $n_s$ and $n_{s\alpha}$ as exposed above, only a slight depletion of the bottom channel is produced.

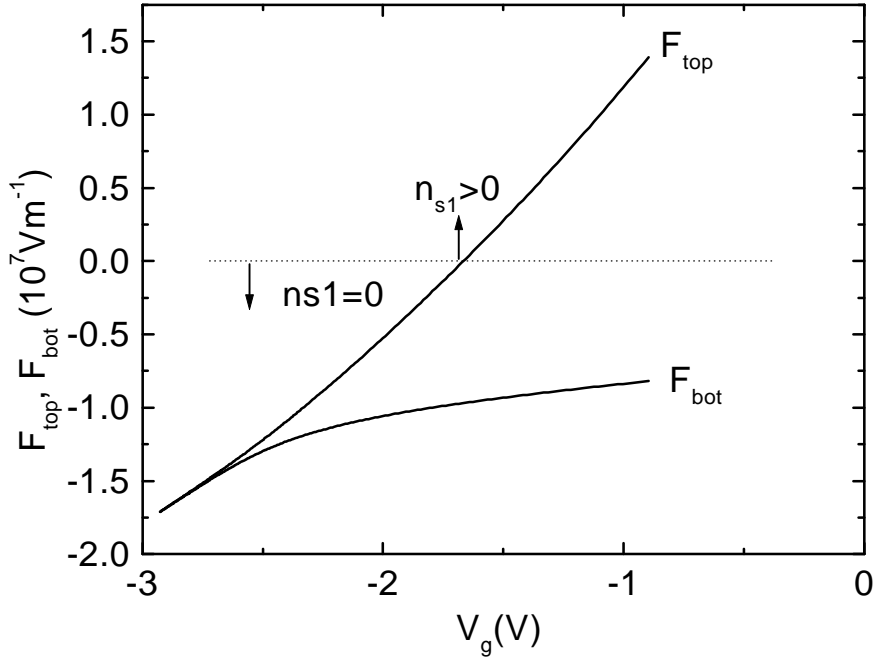

Fig.2  Top and bottom electric fields vs gate voltage, calculated with the analitical model. The parameters are as in [88CAZ]: $N_{D1}=2\times10^{18}$cm$^{-3}$, $N_{D2}=10^{18}$cm$^{-3}$, $d_{i1}=40$ A, $d_{i2}=70$ A, $Y_{1,2}=0.3$, $d_1=440$ A, $d_2=270$ A, $d_w=250$ A, $V_s=1$ V, $V_{F1,F2}=50$ mV.

Under these circumstances, the only argument for the procedure above is a very good agreement seen between numerical and analytical concentrations. The total sheet density almost equals the numerical self-consistent value and, moreover, this remarkable agreement extends over a relatively wide range of parameters ($Y_\alpha$=0.25-0.36, $d_i$=100-450 A, $d_{i\alpha}$=10-100 A, $d_w$=250-350 A) and practically over the whole investigated range of gate voltages $V_g$= -3.5-1 V. In addition, the procedure we propose maintains the continuity of the first derivative $dn_s/dV_g$ at the point where $n_{s1}$ starts increasing. This aspect is of particularly importance for modeling the capacitance-voltage characteristics (Sec.2.4).

iii) To calculate the average distance $Z_{av}$, as a function of $n_s$, we used the following dependence [97CRI]:

$$Z_{av} = (1 - an_s)\sum_i f_i \left( d_w + \frac{2}{3}\gamma_i F_2^{-1/3} \right) \qquad (22)$$

where $a = (1/3.15)\times 10^{-16}$ m$^2$. Except the correction factor $(1-an_s)$, this expression is a *TWA* result and gives the 2DEG average distance from the top interface. The influence of the correction can be understood in the folowing way: the presence of top channel adjust the occupation factors, as naively calculated by TWA, to $(1-an_s)f_i$, and $Z_{av}$ is shifted toward the top interface. Numerical experiments reveal that this trick represents a good compromise with the self-consistent results. As expected, the correction factor plays a significant role at high concentrations, were the top channel is nearly saturated. When the parameter is gradually increased, the maximum predicted sheet density increases while the average distance $Z_{av}$ diminishes (Fig.3). By a proper setting of its value [our choice is $a = (1/3.15) \times 10^{-16}$ m$^2$ throughout in this letter, and is based on self-consistent results] both over and under estimation regions of the sheet density are, to a certain extent, removed.

iv) Introducing $n_s$ and $Z_{av}$ in the charge-control equation (14), we obtain the *2DEG* density as a function of gate voltage. At a certain gate voltage, namely the critical

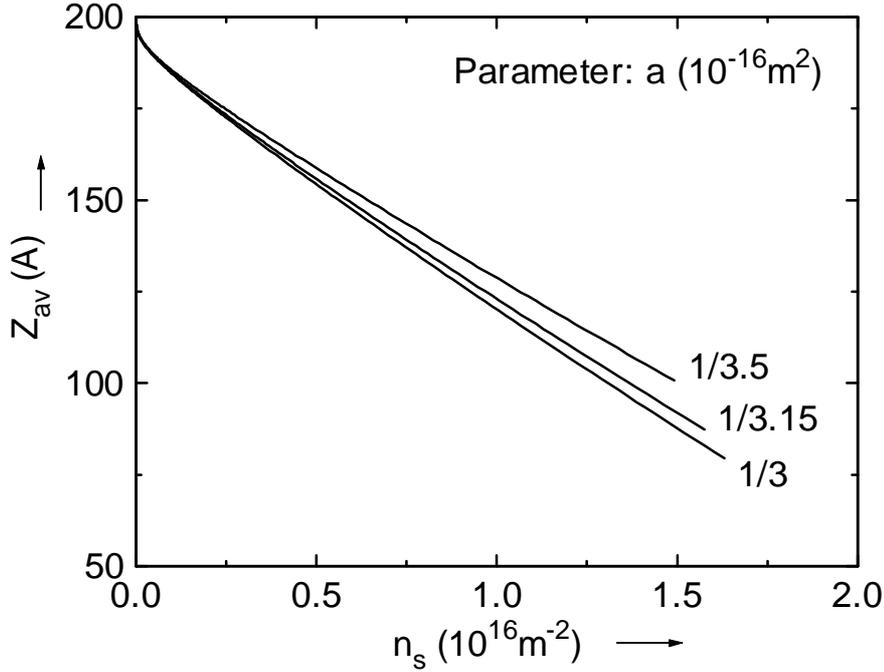

Fig. 3 The average distance from the top interface vs. *2DEG* density, calculated with eq. (22) for $a = 1/3$, $1/3.15$ (used throughout in this letter), and $1/3.5$ ($10^{-16}$ m$^2$). The parameters are as in fig. 2

gate voltage $V_{g,cr}$, the 2DEG density reaches its maximum. Taking into account for the neutrality condition and using (12), at the critical concentration $n_{s,cr}$, the following simple equation holds:

$$n_{s,cr} = N_{D1}(w_1 - d_{i1}) + N_{D2}(w_2 - d_{i2}) \tag{23}$$

Since further increases of the gate voltage can only produce changes in the donor ionization and free electron generation in the top layer, the coupling between the 2DEG-states and the gate voltage appears to be very weak and $n_s$ remain constant. As a result, the top layer is not completely depleted, and some transmission through a MESFET channel between the source and drain is possible [84LEE]. This parasitic conduction leads to a decrease in transconductance and cutoff frequency [88CAZ].

Figure 4 shows how the principle of charge conservation, implied by (23) works. When the critical gate voltage $V_{g,cr}$ is reached, the top depleted thickness $w_1$ must therefore have the same expression as $w_2$, except that, in the latter, the index 1 is to be replaced by index 2. Consequently, the critical concentration is reached when the length $n_{s1}/N_{D1}$ equals the top depleted length $w_1-d_{i1}$. This means that at the critical voltage, in the top AlGaAs layer, $[E_C(-w_1)-E_F]+qV_{FD}= qV_{F1}$. In our case, for a structure with the parameters as in fig. 2, the above equality holds when $w_1 - d_{i1} \cong 50A$

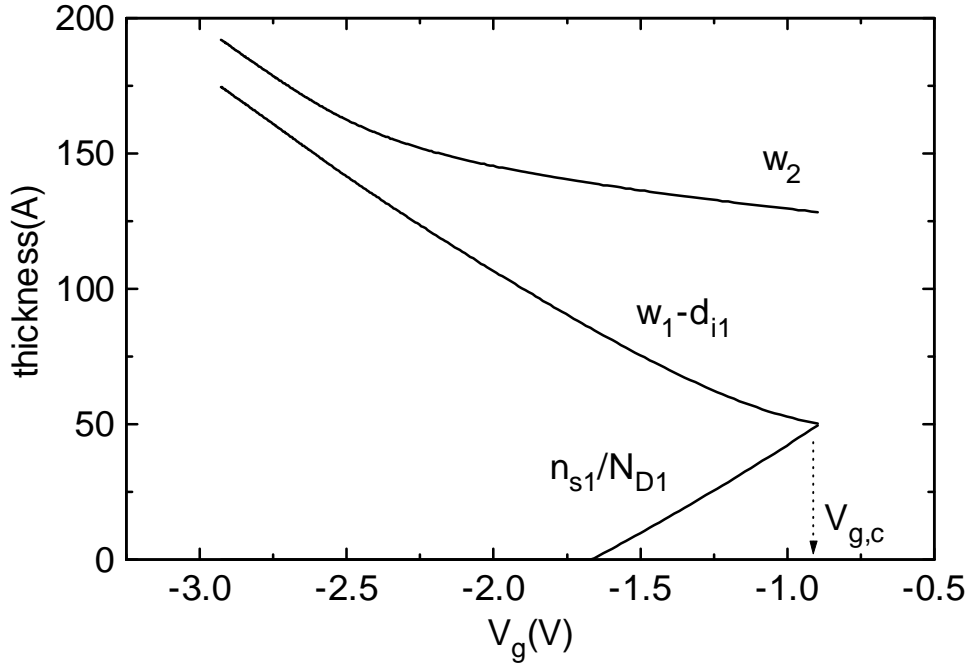

Fig. 4 Depleted thickness vs gate voltage. The parameters are as in fig. 2

A typical dependence of the 2DEG energy spectrum on the gate voltage, calculated with the prescriptions i-iv, is depicted in Fig.5. Only a slight variation in $E_{F2}$ is observed after $n_{s1}$ starts to increase ($V_g = -1.66V$). This happens when $E_{F1}$ intercepts the lowest energetic subband in QW. Such a circumstance leads to a nearly constant sheet concentration in the bottom channel and then maintains the approximations (1a) and (1b). In this moment, at the top interface, $F_1$ vanishes, and a while later enters the positive region where the top chanel is now populated with 2DEG electrons (fig. 2 and fig. 7). Because the flat Fermi level approximations well works, the difference $E_{F1}-E_{F2}$ measures the difference in energy between the bottom of QW at the two interfaces. When $E_{F1}$ intercepts $E_{F2}$ we can speak about a "well balanced" quantum well. Accordingly, the total sheet concentration and the concentration in the top channel are related by (eqs. 1a and 13):

$$n_{s1} = \frac{\varepsilon_1}{\varepsilon} n_s \left(1 - \frac{Z_{av}}{d_w}\right) \tag{24}$$

so that, as long as in such a situation $Z_{av}/d_w$ has the order of 1/2, the top and bottom sheet densities are not very much different.

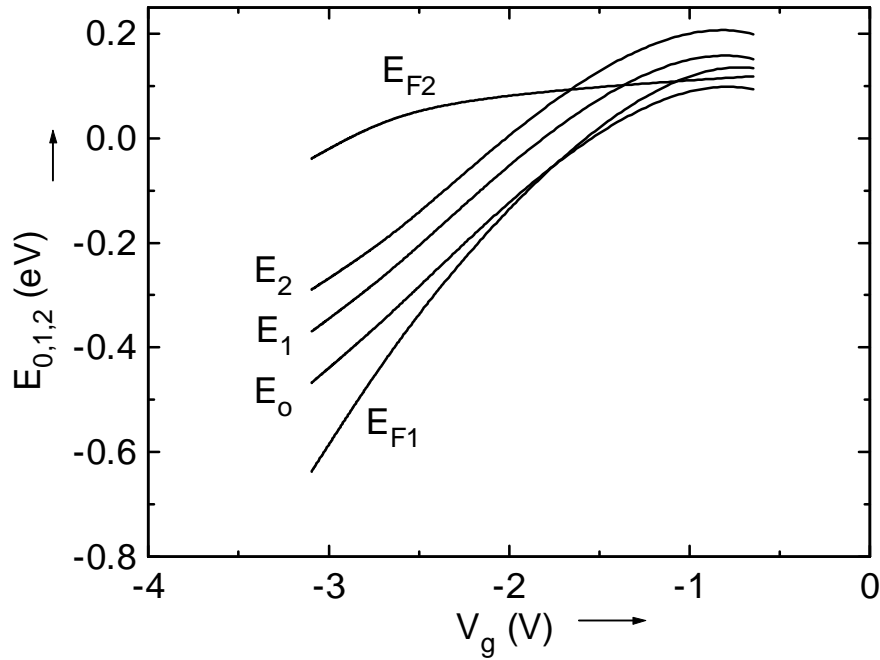

Fig.5 2DEG energy spectrum ($k_y=0$) vs. gate voltage, calculated with the analytical model. The reference is the bottom of the quantum well at the top interface, and $E_{F2}$ is measured from the bottom of the QW at the second interface). The parameters are as in fig. 2

*2.6 Comparisons of the proposed model results to previously self-consistent and analytical models*

We will now perform an analysis of the analytical charge-control model results by comparison with two self-consistent numerical models: one is reported by Inoue *et al.* [85INO] and accounts only for $E_O$, $E_1$ subbands; another one is based on a self-consistent numerical technique [87JAF] that considers five subbands [88CAZ]. First, to calculate the conduction band discontinuity, $\Delta E_{C\alpha}$ and the AlGaAs permitivity, $\varepsilon_\alpha$ (both dependent on Al mole fraction, $Y_\alpha$), we used the same dependence as reported in [85OKU]:

$$\Delta E_{C\alpha} = 0.67\Delta E_g = 0.835 Y_\alpha \quad (25a)$$

$$\varepsilon_\alpha = (12.88-2.82 Y_\alpha)\varepsilon_o \quad (25b)$$

Fig.6 shows the $n_{s\alpha}$ and $n_s$ concentrations versus gate voltage, obtained from our analytical model, and from self-consistent numerical models, respectively. In modeling we used the structure first reported by Inoue *et al.* [85INO]. As discussed in the Sec.2.5, our analytical model predicts that the bottom channel slightly depletes when $n_{s1}$ starts increasing. This behavior has its origin in the particular mechanism we used to generate the total sheet density (as controlled by $E_{F2}$ and TWA approximation), subsequently distributed in the top and bottom channel, by the recipe (21). Anyway, for the total 2DEG density, a very good agreement is seen between the analytical and numerical data, even in the threshold region where $n_S$ is small.

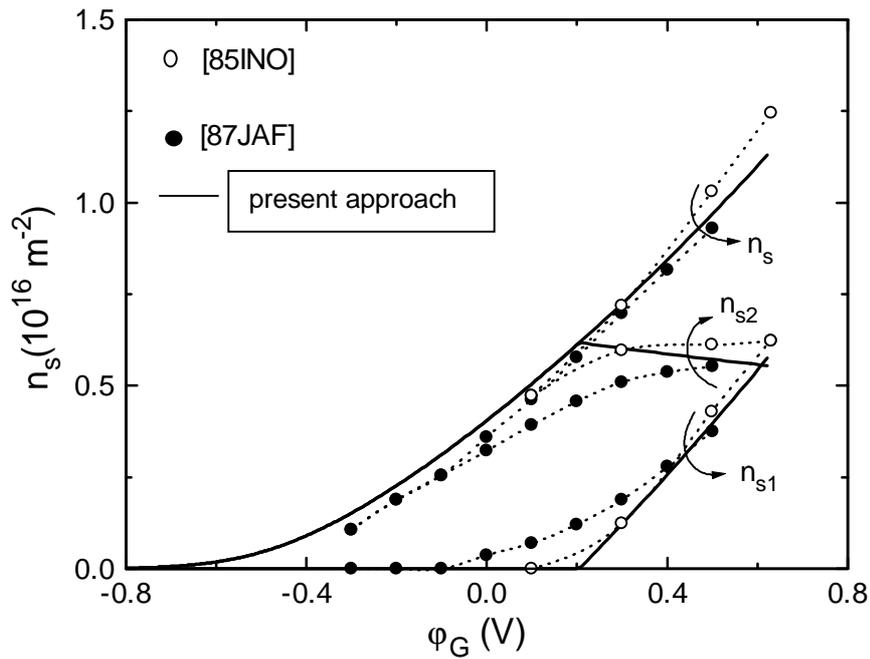

Fig.6 Comparison of the analytical model results (full line) with data from Inoue *et al.*[85INO] (open circles) and numerical solutions using five subbands (solid circles). The parameters are $N_{D1} = N_{D2} = 1 \times 10^{18} cm^{-3}$, $d_{i1} = d_{i2} = 100 A$, $N_A = 0$, $Y_{1,2} = 0.36$, $d_1 = 350 A$, $d_2 = 150 A$, $d_w = 300 A$, $V_S = 1V$, $E_C - E_D = 96 meV$.

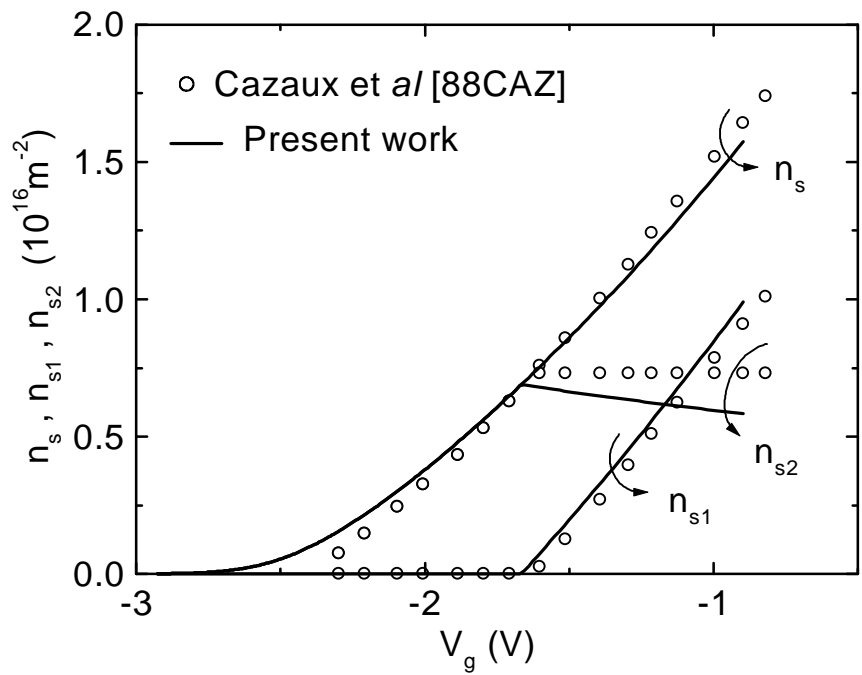

Fig. 7 Comparison with the results of an accurate analytical model reported by Cazaux et al. [88CAZ]

To provide further confirmation for the validity of the present analytical approach, Fig.7 shows a comparison with the results of an accurate analytical model reported by Cazaux *et al.* [88CAZ]. The results from the two models compare well. The largest difference occurs near $V_{g,th}$, where our analytical method predicts a somewhat lower decaying rate of the 2DEG concentration. It should be noted that, in the high voltage range, a lower maximum density is predicted. The measurements of drain currents and Hall measurements of $n_S$, performed on real structures in [88CAZ], seem to confirm these features.

### 3. Influence of physical parameters

The analytical model was used to evaluate the influence of top layer and spacer thicknesses $d_1$, $d_{i\alpha}$, Al composition $Y_\alpha$ and well size $d_w$. Figs.8, 9 show the effect of the spacer layer thickness $d_{i1}$ on $n_s(V_g)$ characteristics.

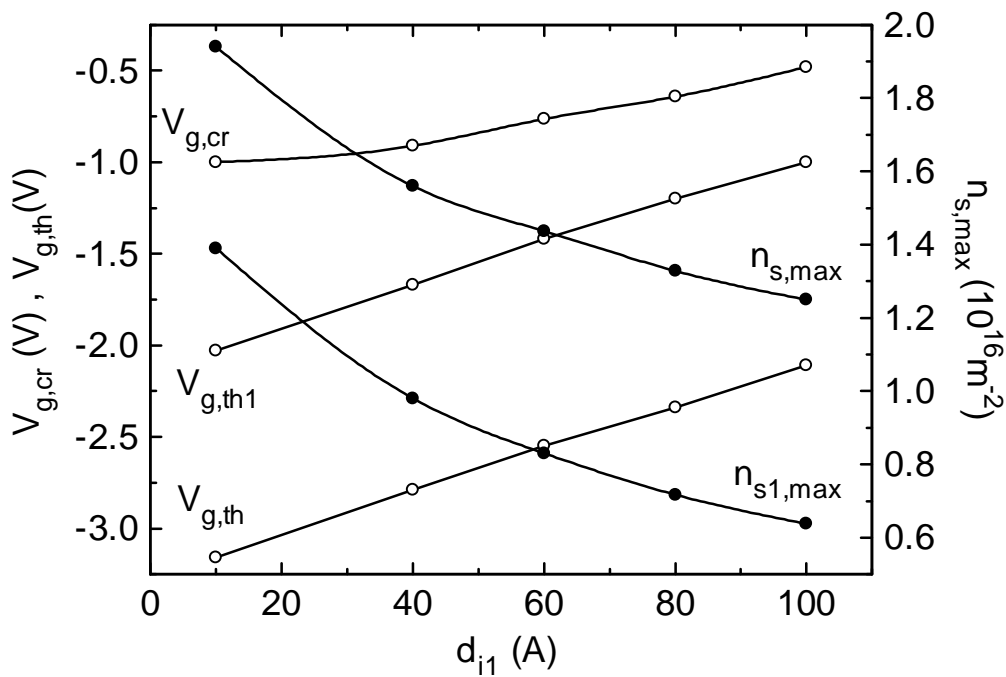

Fig.8 The influence of the top spacer thickness $d_{i1}$ on $n_s(V_g)$ characteristics. Except $d_{i1}$, the parameters are the same as in fig. 2

The maximum 2DEG density has a nonlinear dependence on the top spacer thickness $d_{i1}$. Increasing spacer thickness causes $n_{s,max}$ to decrease while an increase in $V_{g,th1}$, $V_{g,th}$ and $V_{g,c}$ is produced (this means less negative). The $V_{g,th}$ and $V_{g,th1}$ dependencies on $d_{i1}$ are nearly linear and this is a profitable effect that seems to be useful for direct-coupled FET logic DCFL circuits design. Note that $V_{g,cr}$ also depends significantly on spacer thickness. Figs. 11 and 13 show how the capacitance is influenced by the top layer thickness $d_1$. These pictures reveal a nonlinear dependence of $n_s$ on $V_g$ in the whole range of gate voltages. By decreasing $d_1$ this feature becomes more pronounced, until the device enters the normally-off operation domain (fig. 11). As expected, $V_{g,th}$ exhibits a nonlinear dependence on $d_1$ (figs. 11, 12). With the remaining parameters as in fig. 2, an absolute normally-off structure is predicted for $d_1 \sim 100$ A, and a normally-on structure for $d_1 > 300$ A. Interesting enough, the bottom spacer thickness $d_{i2}$ has only little influence on $n_s$-$V_g$ characteristics (fig. 10) and, as a consequence, on $C_g$-$V_g$ dependencies (fig. 15).

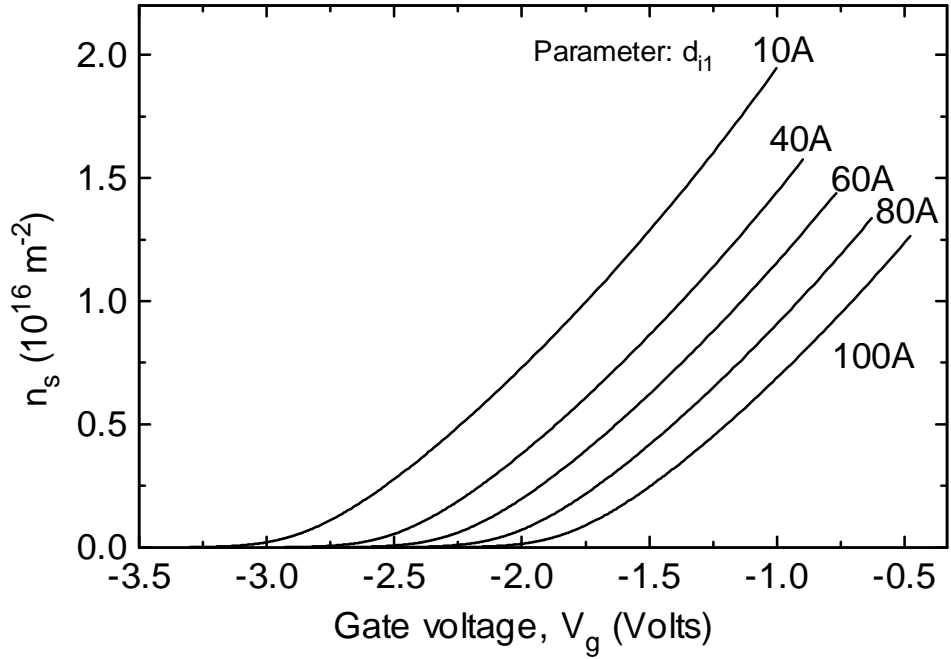

Fig. 9 Total 2DEG concentration as a function of gate voltage. The family parameter is $d_{i1}$ and the remaining structure parameters are as in fig. 2

The effect of $d_{i\alpha}$, $d_w$ and Al mole fractions $Y_2$ on the device capacitance also was investigated (figs. 14-17), and we have obtained similar dependencies as previously reported in the letter of Cazaux *et al* [88CAZ]. Also, as long the concentration of background impurities in QW remains negligible, the gate capacitance is not very much affected by the quantum well thickness (fig. 16).

We have neglected the background impurities in the quantum well region, but this is not mandatory. The model still works at moderate impurities concentrations ($10^{16}$cm$^{-3}$), with the same average separation given by (22), but our choice for the parameter "$a$" in (22) is to be properly altered to obtain accurate predictions of the main performances. However, the new values are very close to our choice in this letter, and all predictions of the model remain essentially unchanged.

On can conclude that a comparison of the theoretical results obtained from our analytical approach and previously reported ones [88CAZ], [85INO], [87JAF], shows that the overall device performance seems to be accurately predicted.

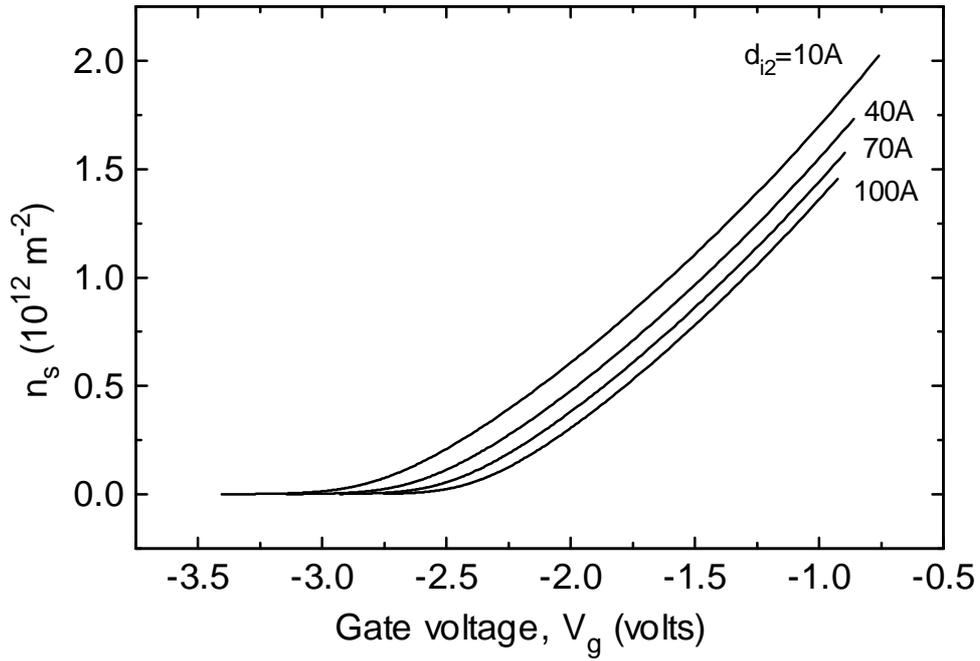

Fig. 10 Total 2DEG concentration as a function of gate voltage. The family parameter is $d_{i2}$ and the remaining structure parameters are as in fig. 2

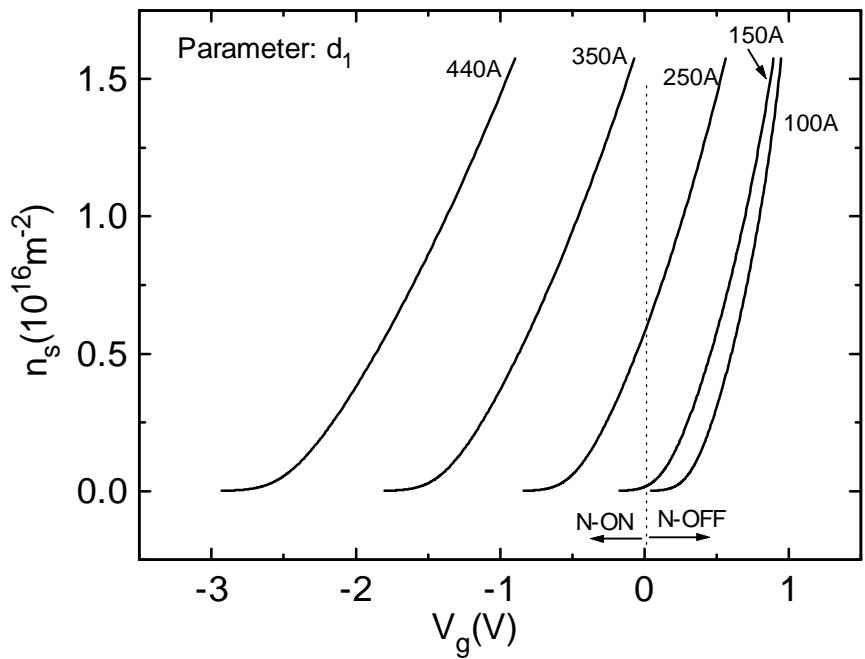

Fig. 11 Total 2DEG concentration as a function of gate voltage. The family parameter is $d_1$ and the remaining structure parameters are as in fig. 2.

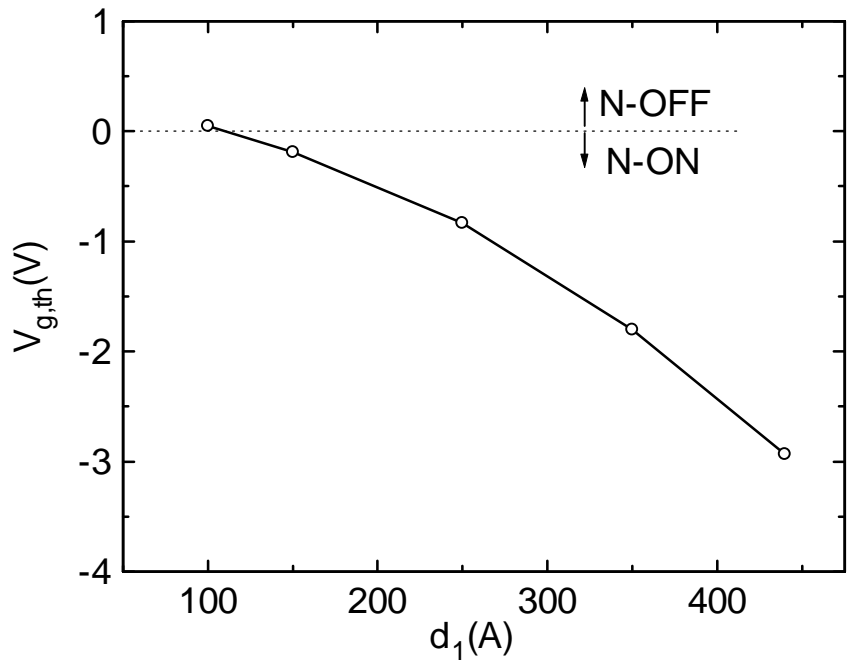

Fig. 12 Threshold voltage as a function of $d_1$. The remaining parameters are as in fig. 2

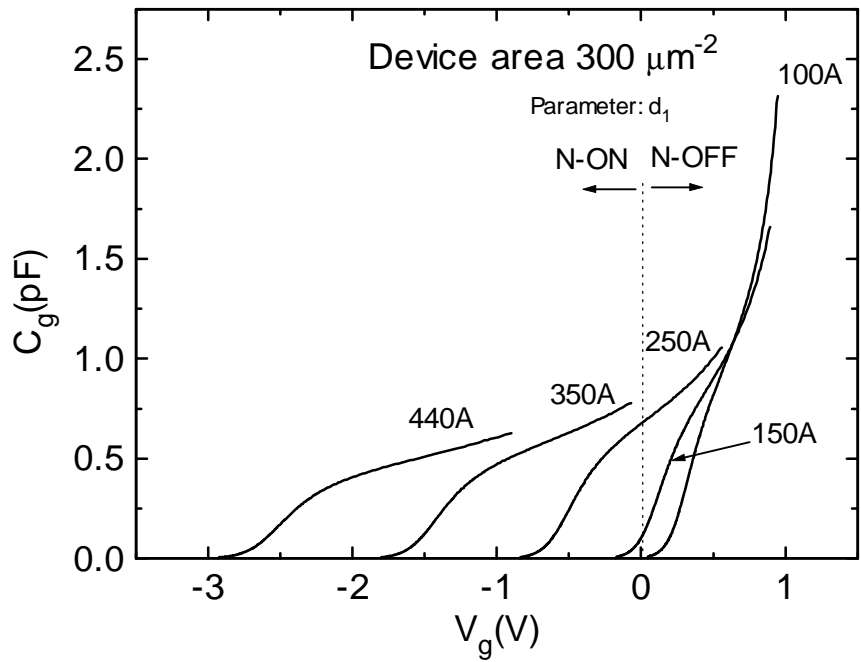

Fig. 13 $C_g$-$V_g$ family of characteristics with $d_1$ as parameter. The remaining parameters are as in fig. 2

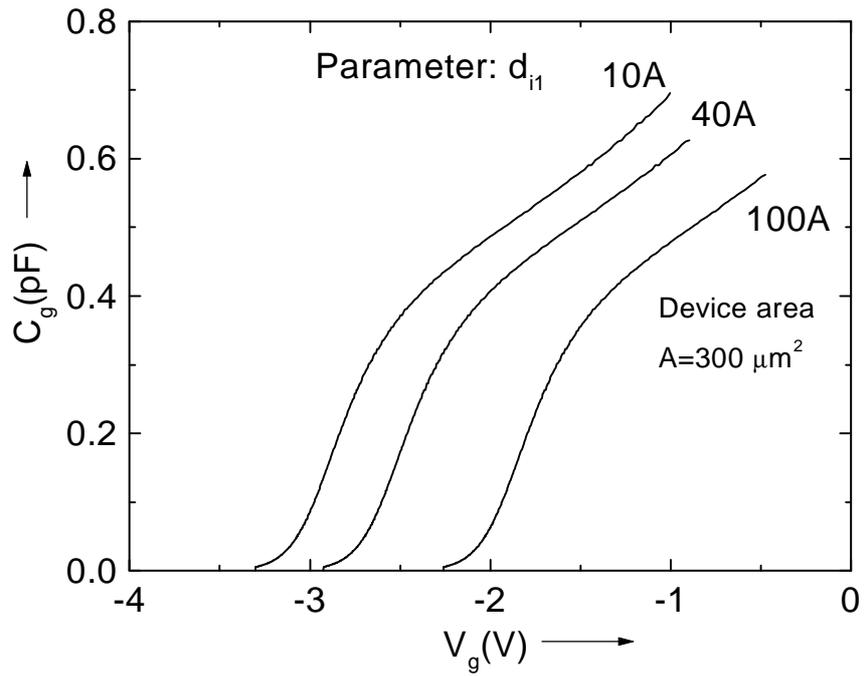

Fig. 14 $C_g$-$V_g$ family of characteristics with $d_{i1}$ as parameter. The remaining parameters are as in fig. 2

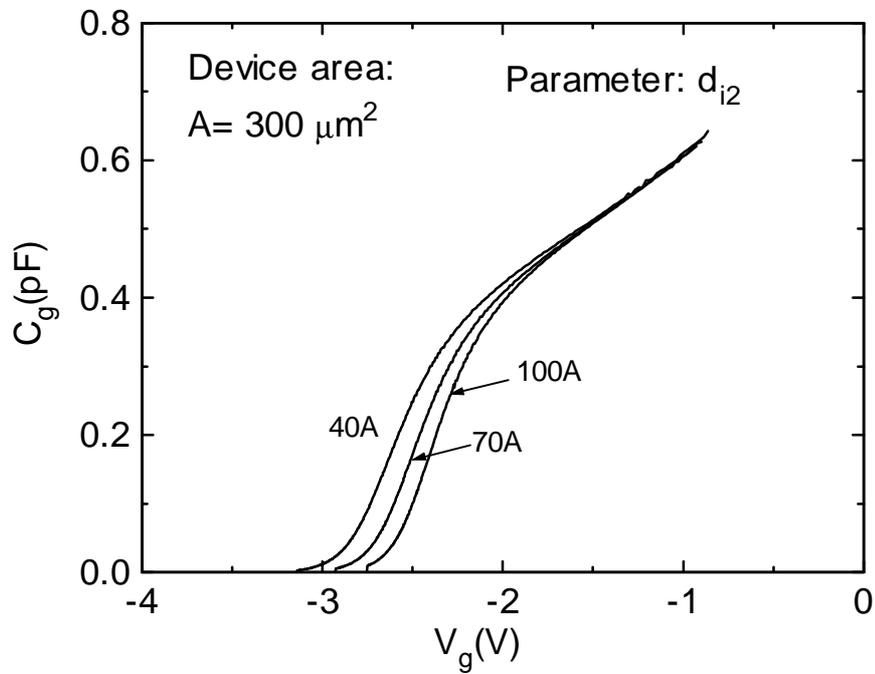

Fig. 15 $C_g$-$V_g$ family of characteristics with $d_{i2}$ as parameter. The remaining parameters are as in fig. 2

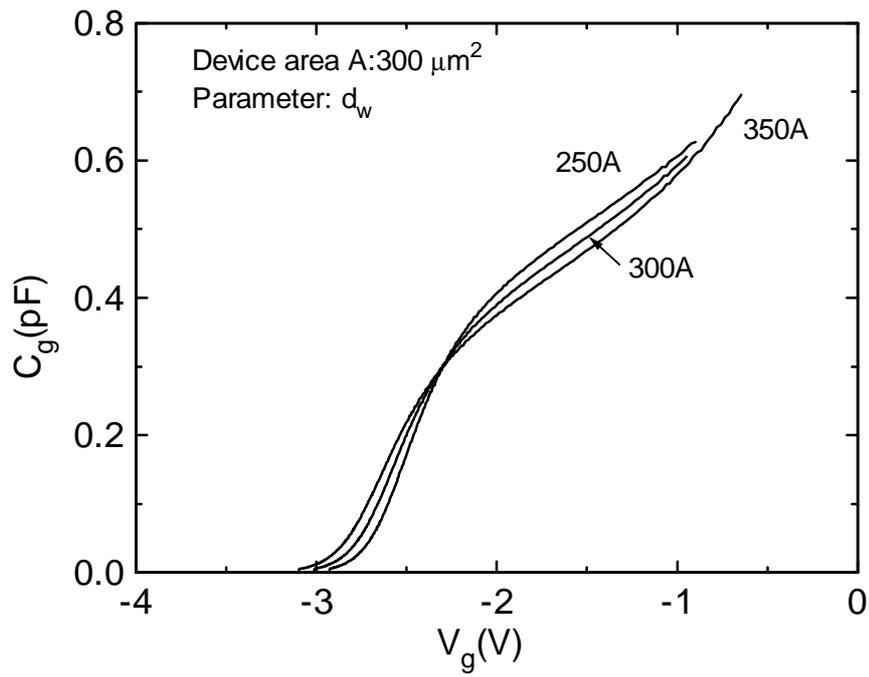

Fig. 16 $C_g$-$V_g$ family of characteristics with the thickness of QW, $d_w$, as parameter. The remaining parameters are as in fig. 2

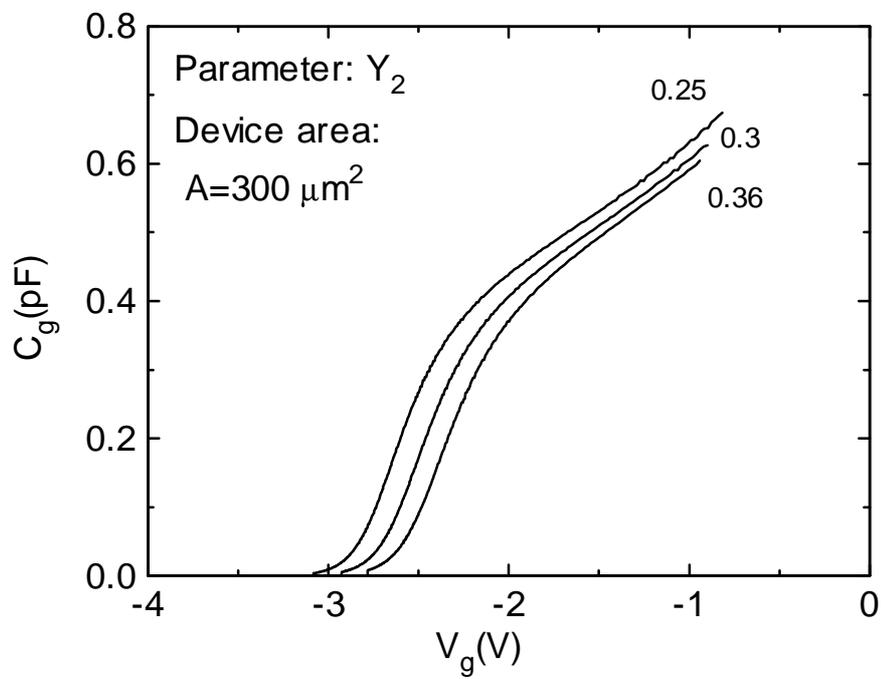

Fig. 17 $C_g$-$V_g$ family of characteristics with $Y_2$ as parameter. The remaining parameters are as in fig. 2

4. Conclusion

A simple analytical approach for charge-control in quantum well region of a DH-HEMT structure was presented. A charge-control equation, accounting for a variable average distance of the 2-DEG from the top interface was introduced. In a simple manner, this equation leads to a general expression for the gate capacitance. Our approach to obtain $n_s$-$V_g$ characteristics is mainly based on a particular mechanism to generate the total sheet density (as controlled by $E_{F2}$ and TWA approximation), subsequently distributed in the top and bottom channel, by the recipe (21). However, the validity of the charge-control equation (14) is not influenced by our particular approach to derive $E_{F2}=E_{F2}(n_s)$ and $Z_{av}(n_s)$.. Such a circumstance is of particularly importance because many others dependencies (generally based on numerical experiments) need to be implemented when studying some nontrivial features of 2DEG dynamics in the QW region. This aspect is to be carefully treated when realistic models of $I_{drain}$-$V_{drain}$ characteristics are desired. However, even with our simple model, by including the effect of source and drain parasitic resistances, and using some crude approximations to derive $I_{drain}$-$V_{drain}$, we have obtained surprisingly good estimations of transconductance and saturation region.

In conclusion, the model we have discussed allows a fast computation and can therefore be used for a realistic estimation of the device performance.

*Appendix I*

If all donors remain fully ionized, by integrating Poisson's equation between $z=-d_1$ and $z=-d_{i1}$, one has:

$$\varphi(z) = -\frac{qN_{D1}}{2\varepsilon_1}z^2 - \left[V_g - V_S + \frac{\Delta E_{C1} - E_{F1}}{q} + \frac{qN_{D1}}{2\varepsilon_1}\left(d_1^2 + d_{i1}^2\right)\right]\frac{z}{d_1} + \frac{E_{F1} - \Delta E_{C1}}{q} - \frac{qN_{D1}}{2\varepsilon_1}d_{i1}^2 \quad (A1)$$

Integrating Poissons's equation between $z=-d_{i1}$ and $z=0^-$ gives:

$$\varphi(z) = -\left[V_g - V_S + \frac{\Delta E_{C1} - E_{F1}}{q} + \frac{qN_{D1}}{2\varepsilon_1}(d_1 - d_{i1})^2\right]\frac{z}{d_1} + \frac{E_{F1} - E_{C1}}{q} \quad (A2)$$

To obtain (A2) we used as boundary conditions $\varphi(-d_1) = V_g - V_S$ and $\varphi(0^-) = (E_{F1} - \Delta E_{C1})/q$, where $qV_S$ is the Schottky barrier height. The reference for the potential energy is the Fermi level and the potential is continuous at $z=-d_{i1}$ (see also Fig.1b of Sec.2.1).

For the bottom layer and $d_{i2} + d_w \leq z \leq d_w + w_2$, using as boundary condition $\varphi(d_w + w_2) = -V_{F2}$ one obtains:

$$\varphi(z) = -\frac{qN_{D2}}{2\varepsilon_2}(d_w + w_2 - z)^2 - V_{F2} \quad (A3)$$

Between $z = d_w$ and $z = d_w + d_{i2}$ we have:

$$\varphi(z) = \frac{qN_{D2}}{2\varepsilon_2}\left[2(z - d_w) - (w_2 + d_{i2})\right](w_2 - d_{i2}) - V_{F2} \tag{A4}$$

The electrostatic potential is continuous at $z = d_w + d_{i2}$ and the boundary condition is $\varphi(d_w^+) = (E_{F2} - \Delta E_{C2})/q$. From (A2) and (A4) we obtain:

$$F_1 = \frac{\varepsilon_1}{\varepsilon d_1}\left[V_g - V_S + \frac{\Delta E_{C1} - E_{F1}}{q} + \frac{qN_{D1}}{2\varepsilon_1}(d_1 - d_{i1})^2\right] \tag{A5}$$

$$F_2 = -\frac{qN_{D2}}{\varepsilon}(w_2 - d_{i2}) \tag{A6}$$

$$w_2 = \sqrt{\frac{2\varepsilon_2}{qN_{D2}}\left(\frac{\Delta E_{C2} - E_{F2}}{q} - V_{F2}\right) + d_{i2}^2} \tag{A7}$$

It is noted that at the top and bottom interfaces we must have $\varepsilon_1 F(0^-) = \varepsilon F_1$ and $\varepsilon F_2 = \varepsilon_2 F(d_w^+)$, respectively. Introducing (10) of Sec.2.2 in (A5) and using (A6), (A7) and (12) of Sec.2.2, rearranging, one obtains the charge-control equation (14) of Sec.2.3 From the above equations, the conduction-band energy diagram for the top and bottom AlGaAs layer can be calculated as a function of gate voltage.

*Appendix II*

From the charge-control equation (14) of Sec.2.3, we first obtain:

$$\frac{dn_s}{dV_g}\left[\left(1 + \frac{\varepsilon_1 Z_{av}}{\varepsilon d_1}\right) + \frac{\varepsilon_1 n_s}{\varepsilon d_1}\frac{dZ_{av}}{dn_s} + \frac{\varepsilon_1}{q^2 d_1}\frac{dE_{F2}}{dn_s} - 2\frac{N_{D2}d_{eff}}{d_1}\frac{d}{dn_s}(d_{eff})\right] = \frac{\varepsilon_1}{qd_1} \tag{B1}$$

We observe that:

$$2d_{eff}\frac{d}{dn_s}(d_{eff}) = -\left[(\varepsilon_1/\varepsilon)d_w + d_1\right]\frac{1}{w_2}\frac{\varepsilon_2}{q^2 N_{D2}}\frac{dE_{F2}}{dn_s} \tag{B2}$$

Introducing (B2) in (B1), one obtains (19) and (19a,b) of Sec.2.4.